\documentclass{article}
\usepackage{setspace}

\usepackage{amsmath} \usepackage{eucal} \usepackage{exscale}

\makeatletter

\newcommand{\LyX}{L\kern-.1667em\lower.25em\hbox{Y}\kern-.125emX\spacefactor1000}

\makeatother

\newcommand{\diag}{\operatorname{diag}}
\newcommand{\const}{\operatorname{const}}

\newcommand{\plabel}{\label}

\begin{document}

\begin{titlepage}
  \renewcommand{\thefootnote}{\fnsymbol{footnote}}

\begin{center}
  
  \hspace*{\fill} TUW-98-16
  
  \vspace*{\fill}

  {\Large Universal conservation law and modified Noether
  symmetry in 2d models of gravity with matter}
        
  \vspace{2ex}
        
  W.~Kummer\footnotemark[1] and G.~Tieber

  \vspace{2ex}
 {\footnotesize Institut f\"ur
    Theoretische Physik \\ Technische Universit\"at Wien \\ Wiedner
    Hauptstra{\ss}e.  8--10, A-1040 Wien, Austria}
             
  \footnotetext[1]{wkummer@tph.tuwien.ac.at}
  
\end{center}

\vspace{2ex}

\begin{abstract}
It is well-known that all 2d models of gravity---including theories
with nonvanishing torsion
and dilaton theories---can be solved exactly, if matter interactions are absent.
An absolutely (in space and time) conserved quantity determines the global classification
of all (classical) solutions. For the special case of spherically reduced Einstein
gravity it coincides with the mass in the Schwarzschild solution. The corresponding
Noether symmetry has been derived previously by P. Widerin and one of the authors
(W.K.) for a specific 2d model with nonvanishing torsion. In the present paper
this is generalized to all covariant 2d theories, including interactions with
matter. The related Noether-like symmetry differs from the usual one. The parameters
for the symmetry transformation of the geometric part and those of the matter
fields are distinct. The total conservation law (a zero-form current) results
from a two stage argument which also involves a consistency condition expressed
by the conservation of a one-form matter ``current''. The black hole
is treated as a special case.
\end{abstract}

\centering{\footnotesize PACS numbers: 04.60.Kz, 04.70.Bw, 11.30.-j}

\vspace*{\fill}\vspace*{\fill}

\renewcommand{\thefootnote}{\arabic{footnote}}
\setcounter{footnote}{0}
\end{titlepage}
\section{Introduction}
\label{Intro}
Models of gravity in one space and one time coordinate have attracted the interest
for some time. After all, spherical reduction of 4d Einstein gravity (SRG) leads
to an effective 2d theory with dilaton fields \cite{TIH84}. Because the Einstein-Hilbert
action becomes trivial in $d=2$ also many other models with or without dilaton
fields have been studied. Ref. \cite{Teit83} contains a necessarily very incomplete
list of related work which had been spurred especially by the discussion
of a special case, the string inspired dilaton black hole \cite{Eli91}. All
these models turned out to be exactly solvable at the classical level in the
absence of matter, for the dilaton black hole of Ref. \cite{Eli91} even (minimal)
interactions with scalar fields may be included. An exact solution is possible
as well for a theory without dilatons but with nonvanishing torsion \cite{Kat86}.
A common behaviour of all these approaches has been the use of the conformal
gauge for the 2d metric. For many nontrivial theories this often required remarkable
mathematical effort to arrive at the exact solution of the equations of motion,
especially in the case with nonvanishing torsion \cite{Kat86}, or for
general dilaton theories \cite{Teit83}.

A new and much simplified approach was first proposed in \cite{KS96}. It is
based upon the consequent use of Cartan variables $e^a_\mu$ (zweibeine) and $\omega^a_{\mu b}= \omega_\mu \epsilon\^a\_b$
(spinconnection) and, more importantly, on a ``light cone gauge'' for those
quantities:
\begin{equation}
\plabel{LCG0}
  e^+_0=(e^-_0-1)= \omega_0=0 
\end{equation}
In that gauge the metric acquires an Eddington-Finkelstein form
\begin{equation}
  \plabel{EF2}
  g_{\alpha \beta}= 2 e^+_1 \left( 
    \begin{array}{cc}
      0&1 \\ 1 & e^-_1
    \end{array}
    \right) \, .
\end{equation}
For the discussion of global properties this gauge has definite advantages,
because---in contrast to the conformal gauge---the basic patch extends across
horizons, the latter being represented by zeros of the Killing norm $k^2=
2 e^+_1 e^-_1$.
This allows a very comprehensive discussion of all global properties of solutions
with one Killing vector field---as it is the case in the absence of matter interactions---in $d=1+1$, and even recently led to the discovery of global coordinates for
the Reissner-Nordstr\"om black hole \cite{KTRN}.

It turns out that the treatment of general 2d theories of gravity is also greatly
facilitated by considering a Hamiltonian formalism \cite{Gro92} or, equivalently,
a first order Lagrangian formalism which has been shown to cover all 2d theories
without dilaton field \cite{TST94}. The solution in that formulation without
matter becomes almost trivial (cf. also \cite{KWid95}) even for a
general gauge, because it essentially
coincides with the solution in component fields, restricted just by the gauge
fixing (\ref{LCG0}). From a theory without dilatons a general dilation theory in $d=2$
always can be produced by conformal transformation of the metric (or
of $e^a_\mu$)
by $\exp \Phi$ where $\Phi$ is a dilaton field. Although this immediately
provides solutions of the equations of motion (e.o.m.) also in that case, the
global properties of the solutions are completely different. The ``decomposition''
of Penrose diagrams in \cite{KKL97} is an illustration for that phenomenon
(cf. also \cite{CM95}).

The exact integrability of all 2d theories in the absence of matter is closely
related to the existence of a conserved quantity which is independent of space
and time. It represents the (only nontrivial) Dirac observable in such systems
\cite{Gro92,TST94} and indeed completely determines the global properties
of the classical solutions for a given 2d action. For the special dilaton theories
describing SRG \cite{TIH84} or the dilaton black hole \cite{Eli91}, not surprisingly,
it is proportional to the mass of the black hole. It may also be interpreted
within the concept of quasilocal energy \cite{KWid95,KL97} or with
a Noether charge \cite{KWid95} in the sense of Ref. \cite{W93}.

According to accepted wisdom a Noether charge should be related to a symmetry
of the action \cite{EN18}. Indeed for the matter free case that symmetry has
been identified by one of the present authors in collaboration with
P. Widerin
\cite{KWid94} for the 2d model of \cite{Kat86}.

Although an exact general solution does not exist anymore---exceptional cases
like the DBH excluded---when matter interactions are present, the general conservation
law still holds. It may be simply derived from a proper linear combination of
the e.o.m.-s \cite{KWid95}, a procedure which is most transparent formulating
2d theories in ``first order form'' \cite{Gro92,KWid95,IK94}. 

However, although the existence of such a conservation law is also
obvious in that case 
\cite{KWid95} the contribution of the matter fields to it would require the
knowledge of at least part of the exact solution coming from the matter interactions.
The precise meaning of that will be recapitulated below. In any case, the composition
of the usual ingredients into the celebrated Noether theorems are changed in
a nontrivial manner.

The purpose of our present work is to clarify this point. Somewhat surprisingly
we obtain a Noether-``like'' situation with subtle modifications of the Noether
theorem.

In section \ref{Gen} we shortly describe the first order form of a general covariant
theory in $1+1$ dimensions and the resulting conservation law. Section \ref{Noe} is devoted
to the discussion of the symmetry, starting in \ref{Matl} with the matterless case.
Here we generalize the result of \cite{KWid94} to arbitrary 2d models of the
type (\ref{vak}) below. Matter interactions are treated at first within a simplified
toy model which has a ``geometric'' and a ``matter'' part and where a complete
exact solution is possible (subsection \ref{SToy}). In that model the main difference
with respect to the ordinary Noether situation can be seen more easily than
in the general case (subsection \ref{Sysf}).

In section \ref{CSRG} we emphasize the importance of our result to the most interesting
application of the general argument, namely the Schwarzschild black hole, interacting
with nonminimal scalar matter. The Noether-like conservation law connects the
mass of the black hole in a highly nontrivial manner with the (incoming and
outgoing) flux of matter.
\section{General 2d action with matter}
\label{Gen}
\subsection{Action of geometric variables}
\label{Act}
The geometric part of the action for all 2d theories can be written as
\begin{equation}
  \plabel{vak}
  L^{(g)}=\int \left( X_a T^a + Xd\omega -\frac {V}{2} \epsilon_{ab} e^a \wedge
  e^b  \right) 
\end{equation}
which depends on the Cartan variables $e^{b}$ (zweibein 1-form) and
$\omega \^a \_b = \epsilon \^a \_b  \omega$
(spin connection one-form) through torsion
\begin{equation}
  \plabel{Ta}
  T^a=de^a+\epsilon^a\_b \omega
  \wedge e^b
\end{equation}
and curvature scalar
\begin{equation}
  -\frac {R}{2}= \star d\omega \, .
\end{equation}
$X^a$ and $X$ are (zero-form) auxiliary fields. They coincide with
conjugate momenta to $e^a_1$, resp. $\omega_1$ when these quantities are
restricted to a space-like surface $x^0=\const$.

In $V=V(X^aX_a,X)$ the dynamics of $L^{(g)}$ are encoded.
Latin indices refer to the tangential (local Lorentz, anholonomic) coordinates
with metric $\eta_{ab}$ [$\diag(\eta)=(1,-1)$]. Greek indices in the components
of the forms ($e^a=e^a_\mu dx^\mu$, $\omega=\omega_\mu dx^\mu$)
indicate space-time (holonomic) ones.

The antisymmetric $\epsilon$ pseudotensor in tangential Minkowski space appearing
in the volume form
\begin{equation}
  \plabel{eps1}
  \epsilon_{ab}= \left(
  \begin{array}{ cc}
   0&1 \\ -1&0
  \end{array}
  \right)
\end{equation}
by
\begin{equation}
  \plabel{eps2}
  \epsilon_{\mu\nu}=\epsilon_{ab} e^a_\mu e^b_\nu = -e \tilde
  \epsilon_{\mu\nu} \, , 
\end{equation}
is related to the antisymmetric symbol $\tilde \epsilon$ in holonomic
components with the factor
\begin{equation}
e= det \, e^a_\mu = \sqrt{-g} \, ,
\end{equation}
the  determinant of the metric
\begin{equation}
\plabel{met} 
  g_{\mu\nu}=e^a_\mu e^b_\nu \eta_{ab} \, .
\end{equation}
Often light cone coordinates are useful. Then (\ref{vak}) and (\ref{Ta}) become
($\eta_{+-} = \eta_{-+}=1$, $\epsilon_{+-}= - \epsilon_{-+}= -1$)
\begin{equation} \plabel{vakLC}
  L=\int \left( X^+ T^- + X^- T^+ + Xd\omega + V(X^+X^-,X) e^+\wedge e^-\right) \end{equation}
and
\begin{equation}
  T^\pm = (d \pm \omega) \wedge e^\pm \, .
\end{equation}
Elimination of $X^a$ and $X$ by their respective
(algebraic) e.o.m.-s from (\ref{vak}), (\ref{vakLC}) clearly produces a covariant 2d theory
with an action which is an arbitrary function in curvature and torsion. However,
it is also possible to ``integrate out'' the components $\omega_\mu$
together with $X^a$. For $V$ quadratic
in $X^a$
\begin{equation}
  \plabel{VU}
  V= \frac {U(X)}{2} X^aX_a+ v(X)
\end{equation}
in this way the action of the most general 2d dilaton theory is produced:
\begin{equation}
  \plabel{CGHSv}
   L=\int dx^2 \sqrt{-g} \left( -\frac{XR}{2}
   + \frac {U(X)}{2} g^{\mu\nu}\partial_\mu X
\partial_\nu X - v(X) \right)  
\end{equation}
Here torsion is zero and the metric has been introduced instead of the zweibeine.
This equivalence between a theory with torsion (\ref{vakLC}) and a general torsionless
dilaton theory (\ref{CGHSv}) had been noted first for the Katanaev-Volovich model \cite{Kat86}
in Ref. \cite{KKL96} which corresponds to the special case $U(X)=\alpha=\const$
and $v=\beta
X^2 + \Lambda$ . The equivalence between (\ref{vak}) with (\ref{VU})
and (\ref{CGHSv}) however even remains true in the quantum version of the theory \cite{KLV97}.
It is crucial for a correct treatment for the global properties of solutions
to (\ref{CGHSv}): An action (\ref{CGHSv})  could also be obtained by conformal transformation of
a torsionless  action [$U=0$ in (\ref{vakLC})]. But then the global properties of those
two theories differ profoundly \cite{KKL97}.

It was known for a long time that 2d theories with an action being a general
function of $R$ and vanishing torsion can be solved exactly \cite{Teit83}.
For a theory quadratic in torsion this holds as well \cite{Kat86}. To show
this in the conformal gauge \cite{Teit83,Kat86} often requires considerable
effort. On the other hand, in the first order form (\ref{vakLC}) this is quite straightforward
\cite{KWid95}. For the potential $V$ in (\ref{VU}) the e.o.m.-s from the action read
\begin{align}
  \plabel{vX-}
  &dX^--X^-\omega+V e^- =0  \, , \\
  \plabel{vX+}
  &dX^++X^+\omega- Ve^+=0  \, , \\
  \plabel{vX3}
  &dX-X^+e^-+X^-e^+=0  \, ,
  \\ 
  \plabel{vT}
  &T^\pm + X^\pm U e^+ \wedge e^-=0  \, , \\
  \plabel{vE3}
  &d\omega + \frac{\partial V}{\partial X} e^+ \wedge e^-=0 \, ,
\end{align}
with the general solution
\begin{align}
  \plabel{VL+}
  &e^+= X^+ e^Q df \, , \\
  \plabel{VL-}
  &e^-= \frac {dX}{X^+} + X^-   e^Q df  \, , \\
  \plabel{VL3}
  &\omega= -\frac{dX^+}{X^+} + V e^Q df \, ,
  \\
  \plabel{VLC}
  &C=C^{(g)} = e^Q X^+X^- + \int_{y_0}^X dy e^{Q(y)} v(y) \, ,
\end{align}
where
\begin{equation}
  \plabel{Q=}
Q= \int_{X_0}^X U(y)dy \, .
\end{equation} 
Eqs. (\ref{VL+})--(\ref{VLC}) depend on arbitrary functions
$X^+$, $f$, $X$ and
one constant $C$. The latter arises by linearly combining (\ref{vX-})--(\ref{vX3}) from
the relation
\begin{equation} 
  \plabel{vakX123}
  d(X^+X^-)+VdX=0 \, .
\end{equation}
Multiplication with the integrating factor $\exp Q$ yields
\begin{equation} 
  \plabel{dCg}
  dC^{(g)}=0 \, .
\end{equation}
Clearly the definition of the constant \( C \) is determined by the chosen
conventions for the lower limits in the integrals: a change of $y_0$ would
redefine $C$ by an additive constant, a change of $X_0$ by a multiplicative
one. For example in SRG ( $U_{SRG}= -(2X)^{-1}$, $v_{SRG}= -4 \lambda^2$)
the choice $X_0=1$, $y_0=0$
yields $C\propto M$, the mass of the black hole (cf.
Section \ref{CSRG}).

Of course, (\ref{vX-})--(\ref{vE3}) also allow various other integrability (consistency)
conditions, which resemble (\ref{dCg}). For example from (\ref{vX-}) and (\ref{vX+}) also
\begin{equation} 
  d(X^\pm \omega - V e^\pm) =0 
\end{equation}
follows. Those integrability conditions are basically different from (\ref{dCg}), because
only (\ref{dCg}) and thus \( C \) alone controls the global properties of the solution
in the metric (\ref{met}). Eliminating $X^+X^-$ through (\ref{VLC}) and
taking $f$, $X$
as coordinates, the metric is found to be expressed in Eddington-Finkelstein
form (\ref{EF2}). $X^+X^- \exp Q$ coincides with
a Killing norm whose zeros (horizons) and behaviour at the (complete or incomplete)
boundary of the interval allowed for $X$ can be used to determine uniquely the
global property of the solution \cite{Kat86,TST94}. Indeed \( C \)
is the only ``observable'' in the sense of Dirac \cite{TST94} and the only
nontrivial element of the center within the (nonlinear, closed) algebra of constraints
and momenta of such a theory \cite{Gro92}.
\subsection{Matter Interactions}
\label{Mat}
Interactions with massless fermions are described by adding to $L^{(g)}$
the action
\begin{equation} \plabel{Ferml1}
  L^{(f)}=  \frac {i}{2}\int  \epsilon_{ab} e^a
  \wedge(\bar \Psi \gamma^b d\Psi - d \bar \Psi\gamma^b\Psi) \, ,
\end{equation}
whereas massless scalars are introduced by
\begin{equation} \plabel{Skl1}
  L^{(S)}= \frac{1}{2} \int F(X) dS \wedge \star
  dS \, .
\end{equation}
The factor $F$ in (\ref{Skl1}) allows the consideration of (nonminimal)
interactions with the dilaton field: for SRG $F=-  \frac{X}{2}$. It is a pecularity of
$d=2$ that not only (\ref{Skl1}) but also (\ref{Ferml1}) do not depend on a spin connection. Therefore
 the e.o.m  from 
 $\delta \omega$ remains unchanged,
whereas (\ref{vX-}) and (\ref{vX+}) now acquire a matter contribution 
\begin{align}
  \plabel{vXM}
  &-\frac{\delta L}{\delta e^\mp} =dX^\pm \pm X^\pm \omega \mp Ve^\pm + W^\pm =0 \, , \\
  \plabel{vX3M}
  &-\frac{\delta L}{\delta \omega}= dX-X^+e^-+X^-e^+=0 \, ,
\end{align}
with
\begin{equation}
\plabel{Wpm}
 W^\pm = \pm F(S^\pm)^2 e^\mp \mp J^\pm \, ,
\end{equation}
where
\begin{equation}
\plabel{Jpm}
J^\mp=i(\chi_{R,L}^\dagger d\chi_{R,L} - (d\chi_{R,L}^\dagger)
\chi_{R,L}) \, .
\end{equation}
$S^\pm$ in (\ref{Wpm}) is an abbreviation for
\begin{equation}
  \plabel{Spm}
  S^\pm= \star dS \wedge e^\pm
\end{equation}
and in (\ref{Jpm}) the two-spinor $\Psi$ has
been expressed in terms of chiral components  as $\Psi=\sqrt [4]{2} (\chi_R, \chi_L)$.

The e.o.m.-s from $\delta X^\pm$, $\delta X$ only receive
a contribution from the matter action $L^{(m)}=L^{(f)}+ L^{(S)}$
in $L=L^{(g)} + L^{(m)}$ if  $\frac{dF}{dX} \ne 0$:  
\begin{align}
  \plabel{ESSE}
  &\frac{\delta L}{\delta X^\mp}=de^\pm \pm \omega \wedge e^\pm +
  X^\pm U e^+ \wedge e^-=0   \\
  \plabel{ESSE3}
  &\frac{\delta L}{\delta X}= d\omega + \left[\frac{dU}{dX} X^+X^- +  \frac{\partial v}{\partial X}
  +\frac{dF}{dX}  S^+S^-  \right] e^+ \wedge e^-=0 
  \end{align}
Eq. (\ref{ESSE3}) exhibits the dependence of the curvature scalar
$R=-2 \star d\omega$,
which is proportional to the square bracket, on nonminimally coupled scalars.
For the fermion field the e.o.m.-s read
\begin{equation}
  \plabel{dLch}
  \pm i \frac{\delta L}{\delta \chi_{R,L}^\dagger}=   2 e^\pm  \wedge
    d\chi_{R,L} -  de^\pm \chi_{R,L} =0 \, ,
\end{equation}
and the same equations for $\chi_{R,L}^\dagger$ ,
whereas for the scalar field
\begin{equation}
  \plabel{dLS}
  - \frac{\delta L}{\delta S}= d(F\star dS) = 0 \, .
\end{equation}
The conservation law generalizing (\ref{dCg}) in the presence of matter is obtained
again by the same linear combination of (\ref{vXM}) with (\ref{vX3M})
\begin{align}
  \plabel{dCM}
  &dC^{(g)} + W^{(m)}=0 \, ,\\
  \plabel{Wm}
  & W^{(m)}= e^Q \left( X^+ W^- + X^-  W^+ \right) \, ,
\end{align}
where $C^{(g)}$ has been defined in (\ref{VLC}). From (\ref{dCM}) the
existence of an absolutely (in space and time) conserved quantity follows \cite{KWid95}.
The integrability condition $dW^{(m)}=0$ within the
conditions of Poincar\'{e}'s lemma implies that the one-form $W^{(m)}$
is
exact, $W^{(m)}=dC^{(m)}$, so that
 \begin{equation}
  \plabel{Cges}
  C=C^{(g)} + C^{(m)} =\const \, .
\end{equation}
That $W^{(m)}$ is closed must be implied as well by th e.o.m.-s which is
indeed the case \cite{T97}. This will be used below. 

No complete analytic solution, comparable to (\ref{VL+})--(\ref{VLC}) for the matterless
case, of (\ref{vXM}), (\ref{vX3M}), (\ref{ESSE})--(\ref{dLS}) is known in the completely general case although
it is possible (in conformal gauge) to reduce the solution to the one of a fourth
order nonlinear partial differential equation for $X$ which (except for the case
of nonminimally coupled scalars, $\frac{dF}{dX} \ne 0$)
is the same in the case with and without matter \cite{T97}.

A general analytic solution is possible if fermions are restricted to be of
one chirality ($\chi_{R}=0$ or $\chi_{L}=0$) \cite{T97,WK92,PT95}, or
if the scalar field is selfdual (or antiselfdual) in the sense $\star
dS = \pm dS$
, i.e. $S^+=0$ or $S^-=0$ \cite{T97,PT95}.
Solutions exist, of course, when $U(X)=0$, $v=\const$   (flat
space) or if $U(X) \ne 0$, $v(X) \ne 0$
 when such a theory is obtained
from a flat theory by conformal transformation. The latter is true for the
string inspired dilaton theory ($V=0$, \cite{Eli91}) and for theories which after
dilatonization asymptotically become Rindler like (\cite{FR}, $V=\const \ne 0$).

However, a partial solution of the matter equations can be found in suitable
gauges. Choosing appropriate components of the matter field as
coordinates \cite{T97,PT95,NA98} $C^{(m)}$ may
be expressible more directly in terms of  component fields. An example for
that will be useful for the interacting Schwarzschild case (SRG) of section
\ref{CSRG}. Consider nonminimally coupled scalars in (\ref{vXM})--(\ref{dLS}). The first integral of (\ref{dLS})---with appropriate assumptions of differentiability---is trivial in terms
of an arbitrary function \( f \):
\begin{equation}
  \plabel{FdS}
  F \star dS= df
\end{equation}
On the other hand, the zero-forms $S^\pm$ in (\ref{Spm}), 
\begin{equation}
  S^\pm= \epsilon^{\mu \nu} (\partial_\mu S) e^\pm_\nu \, , 
\end{equation}
are just the components of $\star dS$ in the basis $e^\pm$:
\begin{align}
  \plabel{dS+-}
  dS&=S^-e^+-S^+e^-  \\
  \plabel{*dS+-}
  \star dS&=S^-e^++S^+e^- 
\end{align}
Solving (\ref{dS+-}) and (\ref{*dS+-}) for $e^\pm$ and using (\ref{FdS})
in (\ref{*dS+-}) yields
\begin{equation}
  \plabel{e+SS}
  e^\pm =\frac{1}{2S^\mp} \left( \frac{df}{F} \pm dS \right) \, .
\end{equation}
With the matter fields $S$ beside $f$ as coordinate, the metric of the line element
\begin{equation}
  \plabel{MeSS}
  ds^2= 2 e^+ \otimes e^-= \frac{1}{2S^+S^-} \left( \frac{df^2}{F^2}-dS^2 \right)
\end{equation}
is diagonal in \( f \) and \( S \). For minimal coupling ($F=1$)
(\ref{MeSS}) is locally conformal.

The matter contribution one-form to the conservation law (\ref{dCg})
in this basis becomes
\begin{equation}
  \plabel{WmS}
  W^{(m)}= \frac{e^Q}{2} \left[ \left( X^-S^+ - X^+S^- \right) df - F
  \left( X^-S^+ + X^+S^- \right) dS \right] = W_f df + W_S dS .
\end{equation}
The consistency condition $dW^{(m)}=0$ guarantees $\partial_S
W^{(m)}_f =\partial_f  W^{(m)}_S$ 
so that (\ref{WmS}) may be formally integrated according to (\ref{dCM})
\begin{equation}
  \plabel{CmS}
  C^{(m)}= \int_{f_0}^f W^{(m)}_f(f',S) df' + \int_{S_0}^S
  W^{(m)}_S(f_0,S')dS' \, , 
\end{equation}
yielding the (inherently nonlocal in the fields) matter part of the (universal) conservation law.

\section{Noether and Noether-like symmetry in $d=2$}
\label{Noe}
In most applications the Noether theory is used to relate a given symmetry to
a conservation law. Here we have to proceed backwards. Clearly the conservation
law (\ref{VLC}) or (\ref{Cges}) can be written as \cite{KWid94}
\begin{equation}
\plabel{sterh}
\partial_\mu J^\mu_\nu=0 \, ,
\end{equation}
where
\begin{equation}
J^\mu_\nu=\delta^\mu_\nu C^{(g)} \, .
\end{equation}
The appearance of an absolutely conserved \( C \) and the possibility to introduce
associated ``currents'' ($J^\mu_0$, $J^\mu_1$ ) is peculiar to $d=2$
\cite{T97}: Invariance of a  general action $L=\int \mathcal{L}$ in $d$
dimensions with respect to $\delta \varphi_a = (\delta ^k \varphi_a)
\wedge \delta
\gamma^k$ requires
that the Lagrangian  transforms as
\begin{equation}
  \plabel{dLFo}
  \delta \mathcal{L}= -dU^k \wedge \delta \gamma^k \, .
\end{equation}
The r.h.s. must be a total divergence, i.e.---in a slight abuse of the form
language---$\delta
\gamma^k=\delta
\gamma^k_{\alpha_1 \cdots
 \alpha_p} dx^{\alpha_1} \wedge \ldots \wedge
  dx^{\alpha_p} $ are assumed to be ``constant
forms'' associated with the infinitesimal global parameters $\delta
\gamma^k_{\alpha_1 \cdots
 \alpha_p}$ of the symmetry. From (\ref{dLFo}) one immediately
derives that on-shell a current form (the factor $(-1)^\kappa$ should take
into account sign changes from commuting forms if $d\varphi_a$ is
moved to the left before dropping it in the derivative)
\begin{equation}
  J^k= U^k + (-1)^\kappa  \frac{\partial
 \mathcal{L}}{\partial d\varphi_a} \wedge \delta^k \varphi_a 
\end{equation}
is closed
\begin{equation}
  dJ^k=0 \, .
\end{equation}
If the $\delta
\gamma^k$ are ``$m$-forms'' then in $d$ dimensions $J^k$ are $(d-m-1)$-forms
related to the usual components $j^k$ of a Noether current by
$j^k=\star J^k$. $j^k$ is covariantly conserved for vanishing torsion.
If the latter is nonzero, still $j^k= e \star J^k$ is conserved in
relation to a partial derivative. 

When the parameters 
are zero-forms---only then the variations themselves are covariant---$J^k$
is a one-form whose components $j^{k \mu}$ coincide with the ones of
the usual Noether current. Here we are interested in $J^k$
becoming zero-forms ($m=d-1$), because then absolutely conserved quantities are
obtained. With fields $\varphi_a$  zero- and one-forms---as for Cartan variables
and matter fields---this can happen only in $d=2$. Thus there is no
obvious generalization  to
 a similar quantity in higher dimensions.

Whereas for the matterless action $L^{(g)}$ this procedure may be
applied directly (subsection \ref{Matl}), for $L=L^{(g)}+ L^{(m)}$ modifications
of the Noether argument will be necessary: A conservation law for the zero-form
current will require the validity of another conservation law for a one-form
current from the matter contributions.
\subsection{Matterless case}
\label{Matl}
Eqs. (\ref{vX-})--(\ref{vX3}) are the result of the variations of $L^{(g)}$
with respect to $-\frac{\delta}{\delta e^\pm}$ and $-\frac{\delta}{\delta \omega}$, respectively. Therefore, their linear
relation (\ref{vakX123}), after multiplication with $\exp Q$ may be written as 
\begin{equation}
  \plabel{dCg=}
  dC^{(g)} = -e^Q  \left( X^+ \frac{\delta }{\delta e^+} + X^-
  \frac{\delta }{\delta e^-} +V \frac{\delta }{\delta \omega} \right)
  L^{(g)} \, .
\end{equation}
The global symmetry transformations with constant infinitesimal
  parameters $ \gamma_\mu$ 
  in a ``one-form'' $\delta \gamma =\delta \gamma_\mu dx^\mu$  can be read
from (\ref{dCg=}):
\begin{align}
  \plabel{SKWe}
  &\delta e^\pm=X^\pm e^Q  \delta \gamma=
  \frac{\partial C^{(g)}}{\partial X^\mp} 
  \delta \gamma \\
  \plabel{SKWo}
  &\delta \omega =V e^Q \delta \gamma =
   \frac{\partial C^{(g)}}{\partial X} 
  \delta \gamma \\
  \plabel{SKX}
  & \delta X^\pm =\delta X= 0
\end{align}
This generalizes the result of \cite{KWid94} to arbitrary (matterless) 2d
theories, including dilaton theories, if the latter are expressed in the same
first order ``torsion'' form [cf. the relation between (\ref{VU}) and (\ref{CGHSv})]. It can
be verified easily that the Lagrangian indeed varies to a total divergence
\begin{equation}
  \plabel{VLG}
  \delta \mathcal{L}=d (2X^+X^-+XV-C^{(g)}) \wedge \delta \gamma \, ,
\end{equation}
an exact ``form'' in the sense of (\ref{dLFo}).

Comparing (\ref{SKWe})--(\ref{SKX}) with the analytic solution
(\ref{VL+})--(\ref{VL3}) the symmetry is seen
to correspond to an ``orbit'' with $df \rightarrow \delta \gamma$,
$dX^\pm \rightarrow \delta X^\pm=0$, $dX\rightarrow \delta X =0$.
This symmetry commutes with the gauge-symmetries of the theory (local Lorentz
transformations and diffeomorphisms). The last form of (\ref{SKWe}) and (\ref{SKWo}) emphasizes
the close relation to the expression of $C^{(g)}$, interpreted
as a quasilocal energy in a Hamiltonian, employing the Regge-Teitlboim trick
for a boundary term \cite{KWid95}.

There are many more global symmetries which can be related to integrability
conditions follwing from the e.o.m.-s \cite{T97}. They simply reflect the large
gauge freedom and include (not unexpectedly) global Lorentz transformations
and global translations (conservation of energy momentum tensor). It should
be noted though, that especially the latter are---for a nontrivial $d=2$ model---not directly related to the conservation law discussed here \cite{KWid95}.
\subsection{Symmetry with matter: A toy model}
\label{SToy}
As emphasized in subsection \ref{Mat} the conservation law with matter generalizes to eq
(\ref{dCM}). $dC^{(g)} + W^{(m)}$  may be expressed by the same e.o.m.-s
as in (\ref{dCg=}). For   $W^{(m)}$
its integrability condition $dW^{(m)}$ may be related linearly
to the  e.o.m.-s (\ref{vXM}), (\ref{vX3M}), (\ref{ESSE})--(\ref{dLS}). On that basis \emph{another} ``matter''
symmetry transformation may be defined. It is unusual that the Noether symmetry
thus must be ``extended'' when a new piece (here the matter interaction) is
added to the action.

We elucidate this point first for a (topological) toy model in $d=2$ whose action \begin{equation}
  \plabel{toyL}
  L= \int (X dw +Kw \wedge d\varphi) 
\end{equation}
depends on zero-form fields $X$, $K$  and \( \varphi  \) and
on one 1-form field \( w \). The first term of (\ref{toyL}) is a simplified ``geometric''
action whereas in the second term ``matter'', described by an ``amplitude''
\( K \) and a ``phase'' \( \varphi  \) is coupled to the ``geometric''
variable \( w \). Thus this term is made to resemble the fermionic interaction
(\ref{Ferml1}).

The e.o.m.-s for (\ref{toyL}) read
\begin{align}
  \plabel{toyX}
  & dX+K d\varphi =0  \, , \\
  \plabel{toyS}
  &dw =0 \, , \\
  \plabel{toyph}
  &w \wedge d\varphi=0 \, , \\
  \plabel{toyK}
  &d(Kw) =0 \, .
\end{align}
Comparing (\ref{toyX}) with (\ref{dCM}) one notices that (\ref{toyX}) can be interpreted as the counterpart
of a ''conservation law'' with matter if $K \ne 0$. \( X \) takes
the role of $C^{(g)}$. Applying an exterior derivative to (\ref{toyX}) leads
to the integrability condition 
\begin{equation}
  \plabel{toykon}
  dK \wedge d\varphi =0 
\end{equation}
which implies $K=K(\varphi)$ so that
\begin{equation}
  K d\varphi = d \left( \int_{y_0}^\varphi K(y) dy \right)
\end{equation}
and
\begin{equation}
  \plabel{toydC}
  d \left( X + \int_{y_0}^\varphi K(y) dy \right)= 0 \, .
\end{equation}
For a fixed choice of $y_0$ different values of the constant $C$ in 
\begin{equation}
  \plabel{toyC}
  C = X + \int_{y_0}^\varphi K(y) dy 
\end{equation}
characterize the solutions.

From (\ref{toyS}) and (\ref{toyph}) in an analogous way $w=w(\varphi)$
 may be concluded, (\ref{toyK})
 is fulfilled identically. Thus the action (\ref{toyL}) on-shell yields a one-dimensional
theory in which \( \varphi  \) may be considered the only independent coordinate.

In the ``matterless'' case (\( K=0 \)) the (``geometric'') symmetry transformations,
belonging to $-\frac{\delta L}{\delta w}=dX=0$ are translations with the constant ``one-form''
$\delta \gamma$
\begin{equation}
  \plabel{toysym0}
  \delta w= \delta \gamma  \, .
\end{equation}
For \( K\neq 0 \) the integrability condition (\ref{toykon}) allows an expansion in terms
of Eqs. (\ref{toyS}), (\ref{toyph}), (\ref{toyK}), i.e. $\frac{\delta
  L}{\delta X}$, $\frac{\delta
  L}{\delta K}$, $\frac{\delta
  L}{\delta \varphi}$:
\begin{equation}
   \plabel{toyLinkon}
  dK \wedge d\varphi =\left( \frac{\delta L}{\delta \varphi} - K
  \frac{\delta L}{\delta X} \right)  \frac{\partial_0 \varphi}{w_0} +
  \frac{\delta L}{\delta K} \frac{\partial_0 K}{w_0}
\end{equation}
The apparent dependence on the zero components of \( w \) and on
$\frac{\partial}{\partial x^0}$
is spurious. Actually the same equation holds with zero replaced by one. In
fact,
\begin{equation}
  \frac{\partial_0 \varphi}{w_0}= \frac{\partial_1 \varphi}{w_1}
\end{equation}
is nothing else but (\ref{toyph}) written in components. Thus both
formulations on-shell are equivalent.
Eq. (\ref{toyLinkon}) allows the introduction of a (``matter'') symmetry with global (zero-form) parameter $\delta \rho$:
\begin{align}
  \plabel{toykph}
  &\delta \varphi= \frac{\partial_0 \varphi}{w_0} \delta \rho  \\
  &\delta X=- K \frac{\partial_0 \varphi}{w_0} \delta \rho  \\
  \plabel{toykK}
  &\delta K= \frac{\partial_0 K}{w_0} \delta \rho 
\end{align}
or a similar one with $\partial_0 \rightarrow \partial_1$, $w_0
\rightarrow w_1$. The Lagrangian $\hat \mathcal{L}$ in $L= \int \hat
\mathcal{L}$  under (\ref{toykph})--(\ref{toykK}) transforms as a
total divergence
\begin{equation}
  \delta \hat \mathcal{L}=d  \left(  K d\varphi - K  \frac{\partial_0
  \varphi}{w_0} w 
  \right)  \delta \rho
\end{equation}
and the related conserved Noether current (one-form) is
\begin{equation}
  J=K d\varphi \, ,
\end{equation}
or in components  of $\star J$ (cf. subsection \ref{Matl})
\begin{equation}
  j^\mu=\epsilon^{\mu \nu} K \partial_\nu \varphi \, .
\end{equation}
The conservation $dJ=0$ of the one-form current just yields the
integrability condition, i.e.
that the zero-form conservation law from (\ref{toyX}) can be written
as (\ref{toydC}) or (\ref{toyC}).
We thus observe a ``two stage'' argument instead of the usual direct one with
one single symmetry transformation for all fields. Also the result of $dJ=0$, 
i.e. $K=K(\varphi)$ is nonstandard. It means that (part of) the conserved
quantities are not simply expressed by a local combination of the fields appearing
in the action, but that those fields may become functions of each other.
\subsection{Symmetry with scalars and fermions}
\label{Sysf}
Eq. (\ref{dCg=}) from (\ref{dCM}) now is replaced by
\begin{equation}
  \plabel{dCg+=}
  dC^{(g)} + W^{(m)} = -e^Q  \left( X^+ \frac{\delta }{\delta e^+} + X^-
  \frac{\delta }{\delta e^-} +V \frac{\delta }{\delta \omega} \right)
  L \, ,
\end{equation}
where the e.o.m.-s (\ref{vXM}) and (\ref{vX3M}) appear on the r.h.s. Introducing the same (``geometric'')
global parameters with the associated transformation law (\ref{SKWe})--(\ref{SKX}) to be read
off again from (\ref{dCg+=}) as in the toy model one does not arrive
at the complete conservation law,
but only at (\ref{dCM}). The ``secondary'' conservation law for $W^{(m)}$
is expressed in terms of th e.o.m.-s (\ref{vXM}), (\ref{vX3M}),
(\ref{ESSE}), (\ref{dLch}), (\ref{dLS}). This is straightforward even
without gauge fixing. But in order to simplify the resulting equation we choose
a special set of conformal coordinates. 

The fermion currents $J^\pm$ in (\ref{Wpm}) may be written as function
of ``amplitude'' and ``phase'' as \cite{PT95}
 \begin{equation}
\chi_{R,L} = \frac {1}{\sqrt{2}} k_{R,L} e^{i\varphi^{R,L}} \, .
\end{equation}
For $K^\pm= k^2_{L,R}$ this yields
\begin{equation}
  J^\mp = - K^\mp d\varphi^{R,L} \, .
\end{equation}
A conformal gauge results by identifying $d\varphi^{R}$,
$d\varphi^{L}$ with the coordinates.
For a one-form we thus have $\omega = \omega_R d\varphi^{R} +\omega_Ld\varphi^{L}$   etc. These coordinates
may be used---within the conformal patch---also when no fermions are present
($K^\pm=0$). Of course, the relation to the space-time coordinates may
become highly nontrivial. In terms of $d\varphi^{R,L}$ the e.o.m.-s are unchanged,
except (\ref{dLch}) which is replaced by
\begin{align}
  \plabel{EFG1}
  &\frac{\delta L}{\delta K^+} =-e^-\wedge d\varphi_L=0 \, ,\\
  \plabel{EFG2}
  &\frac{\delta L}{\delta K^-} = e^+\wedge d\varphi_R=0 \, ,\\
  \plabel{EFG3}
  &\frac{\delta L}{\delta \varphi_R} =d(K^-e^+)=0 \, , \\
  \plabel{EFG4}
  &\frac{\delta L}{\delta \varphi_L} = -d(K^+e^-)=0  \, . 
\end{align}
The line element from (\ref{EFG1}) and (\ref{EFG2})
\begin{equation}
  (ds)^2= 2 e^+ \otimes e^- =  2 e^+_R e^-_L d\varphi^R d\varphi^L
\end{equation}
indeed is conformal ($e^+_L = e^-_R = 0$). The one-form conservation
law for matter $dW^{(m)}$ becomes a linear combination
of (\ref{vXM}), (\ref{vX3M}), (\ref{ESSE}), (\ref{dLS}),
(\ref{EFG1})--(\ref{EFG4}):
\begin{equation}
  \begin{split}
   d W^{(m)} 
   &= e^Q \left[ \frac{\partial_0
   \varphi^R}{ e^+_0}   \left( X^+  \frac{\delta }{\delta \varphi^R}
   - X^+ K^- \frac{\delta
   }{\delta X^-}  +  K^-  e^+ \wedge \frac{\delta
   }{\delta e^-}   \right)   \right. \\
   &\left. - \frac{K^-}{e^+_0} \left( K^+ \partial_0
   \varphi^L - \frac{\partial_0 K^-}{K^-} X^+ - \partial_0 X^+ +
   F S^{+2} e^-_0 + U X^{+2}   e^-_0
   \right)  \frac{\delta }{\delta K^-}     \right. \\
   &\left. + \frac{\partial_0
   \varphi^L}{ e^-_0} \left( X^- \frac{\delta }{\delta \varphi^L}   +
   X^- K^+ \frac{\delta
   }{\delta X^+}  - K^+ e^-  \wedge  \frac{\delta
   }{\delta e^+}   \right)   \right. \\
   &\left. +  \frac{K^+}{e^-_0} \left( K^- \partial_0
   \varphi^R + \frac{\partial_0 K^+}{K^+} X^- + \partial_0 X^- -
    F S^{-2} e^+_0 - U X^{-2}    e^+_0
   \right) \frac{\delta }{\delta K^+}    \right. \\
   &\left. + F S^{+2} e^- \wedge \frac{\delta
   }{\delta e^+}    - FX^- S^{+2} \frac{\delta
   }{\delta X^+}  -  F S^{-2} e^+ \wedge \frac{\delta
   }{\delta e^-}  +  FX^+ S^{-2}  \frac{\delta
  }{\delta X^-}  \right. \\ 
   &\ \left. + \left( - F'S^+S^-(X^-e^+-X^+e^-) + U X^+ K^- d\varphi^R
   - U
   X^- K^+ d\varphi^L  \right. \right. \\
   &\left. \left. + UF \left( S^{+2} X^- e^- - S^{-2} X^+ e^+
   \right) \right) \wedge \frac{\delta }{\delta
  \omega}    \right.\\
   &\left. - (X^+S^- - X^-S^+ )  \frac{\delta }{\delta S} + U X^+X^-
   \left( K^+ \frac{\delta }{\delta K^+} - K^- \frac{\delta }{\delta
   K^-} \right)
    \right]L \, .
   \end{split}
\end{equation}
This corresponds to a symmetry with ``matter'' parameter $\delta \rho$
\allowdisplaybreaks
\begin{align}
  \plabel{SySe+}
  &\delta e^+=\left( K^+ \frac{\partial_0 \varphi^L}{ e^-_0}  - FS^{+2}   \right) e^Q e^- \delta \rho \, , \\
  &\delta e^-= \left( - K^- \frac{\partial_0 \varphi^R}{ e^+_0}  + FS^{-2}   \right) e^Q  e^+ \delta \rho \, , \\  \notag
  &\delta \omega = \left[ F'S^+S^- \left( X^-e^+-X^+e^- \right) -  U (X^+ K^- d\varphi^R
   - 
   X^- K^+ d\varphi^L)  \right. \\ 
  &\left. \qquad \quad- UF \left(S^{+2} X^- e^- - S^{-2} X^+ e^+ \right)
   \right] e^Q  \delta \rho \, ,
  \\ \notag \\
  &\delta X^+= \left( K^+ \frac{\partial_0 \varphi^L}{ e^-_0}  - FS^{+2}   \right) e^Q X^- \delta \rho \, , \\
  &\delta X^-= \left( - K^- \frac{\partial_0 \varphi^R}{ e^+_0}  + FS^{-2}   \right) e^Q  X^+ \delta \rho   \, , \\
  \notag \\ \notag 
  &\delta  K^+= \Biggl[ \left( K^- \partial_0
   \varphi^R + \frac{\partial_0 K^+}{K^+} X^- + \partial_0 X^-
   -FS^{-2} e^+_0 - U X^{-2}  e^+_0
   \right)   \frac{K^+}{e^-_0}  \Biggr. \\ 
  &\Biggl.  \qquad \qquad +U X^+X^- K^+ \Biggr] e^Q  \delta \rho \, , \\ \notag
  &\delta  K^-= \Biggl[ \left(- K^+ \partial_0
   \varphi^L + \frac{\partial_0 K^-}{K^-} X^+ + \partial_0 X^+ +
   FS^{+2} e^ -_0 +  U X^{+2}  e^-_0
   \right)  \frac{K^-}{e^+_0}    \Biggr. \\
  &\Biggl. \qquad \qquad - U X^+X^- K^- \Biggr] e^Q \delta \rho \, , \\
  &\delta \varphi_{R}= X^+  \frac{\partial_0 \varphi^R}{ e^+_0} e^Q 
  \delta \rho \, , \\
  &\delta \varphi_{L}= X^- \frac{\partial_0
   \varphi^L}{ e^-_0}  e^Q \delta \rho \, , \\
  \notag \\
  \plabel{SySS}
  &\delta S=  \left (X^-S^+ - X^+S^- \right) e^Q \delta \rho \, . 
\end{align}
Of course the use of $\varphi^{R,L}$  as coordinates is only one way to write
that symmetry.
\section{Conservation law and symmetry for spherically reduced gravity}
\label{CSRG}
As mentioned above, SRG is contained as a special case in our general formalism:
with
\begin{equation}
  V_{SRG}= -\frac{1}{2X} X^+X^- - 4\lambda^2
\end{equation}
for (\ref{VU}) we obtain from (\ref{VLC}) and (\ref{Q=}) for the geometric part of the conservation
law ($x_0=1$, $y_0=0$)
\begin{equation}
  C^{(g)}_{SRG}= \frac{1}{\sqrt{X}} X^+X^- - 8\lambda^2 \sqrt{X} \,
  .
\end{equation}
When the influx of matter is absent (for example after the black hole has been
created in such a way that no spherical wave bounces back), $C^{(g)}$
by itself is constant. The line element from (\ref{VL+}) and (\ref{VL-})
\begin{equation}
  \plabel{metSRG}
  (ds)^2 = \frac{2}{\sqrt{X}} df \otimes \left[ dX +  \left(C +8 \lambda^2 \sqrt{X}
  \right) df \right]  = d\tilde f \otimes \left[ 2 d\tilde X +  
      \left( \frac{C}{16 \lambda^3\tilde X} +1  \right) d\tilde
  f \right]
\end{equation}
in its second version with $\tilde X=\frac{\sqrt{X}}{2\lambda}$,
$f=\frac{\tilde f}{4\lambda}$ exhibits
the Eddington-Finkelstein form (\ref{EF2}). Requiring a diagonal metric in terms of
new variables \( r \) and \( t \) instead of $\tilde X$  and $\tilde f$
with $\det g =-1$  fixes $\tilde X \propto r$
so that indeed $C \propto M$, the mass of the black
hole. The symmetry (\ref{SKWe}) to (\ref{SKX}) may be simply related \cite{KWid95} to a translation
in the direction of the Killing field $\frac{\partial}{\partial f}
\propto\frac{\partial}{\partial t}$ for the metric (\ref{metSRG}). The connection of such a translation with the (conserved) mass of the
black hole is well-known.

The more interesting case arises, when matter interactions are present (e.g.
during the time of formation and also if the formation is accompanied by an
outgoing matter wave \cite{GR97}).

For nonminimally coupled scalars we have to set  $F_{SRG}=- \frac{X}{2}$
in
(\ref{WmS}):
\begin{equation}
  \plabel{WmSRG}
  W^{(m)}_{SRG}= \frac{1}{2 \sqrt{X}}  \left( X^-S^+ - X^+S^- \right)
  df + \frac{\sqrt{X}}{4}
  \left( X^-S^+ + X^+S^- \right) dS
\end{equation}
Then $C^{(m)}_{SRG}$ is the appropriate special case of (\ref{CmS}).

The line-element (\ref{MeSS})  for $F_{SRG}=- \frac{X}{2}$ corresponds to the choice of
\( f\) and \( S \) as ``coordinates'' but by a suitable choice of \( f =f(t,r)\), \( S= S(t,r) \),
 an Eddington -Finkelstein metric and with $C$
replaced by a function of \( r \) and \( t \) (``variable mass'') may be
obtained. This clearly is possible only where the matter-field does not vanish
identically. Therefore (\ref{WmSRG}) may be useful when SRG behaviour is described ``inside''
the 2d submanifold, occupied by the developping matter field. Of course, the
introduction of $dS$ as variable is not mandatory. An Eddington-Finkelstein
gauge choice [cf. (\ref{EF2})] with $e^+_0=(e^-_0-1)= (e^+_1-1) =0$,
$e^-_1 = e^-_1 (t,r)$ 
would have the advantage of being extendable across the horizons
(zeros of $e^-_1$).
But---in contrast to the matterless case---that $e^-_1$ would have
no immediate relation any more to the conserved quantity $C$.

The symmetry relation from (\ref{WmSRG}) is contained in
(\ref{SySe+})--(\ref{SySS})
:
\begin{align}
  &\delta e^+=    \frac{\sqrt{X}}{2} S^{+2} e^-  \delta \rho  \\
  &\delta e^-= - \frac{\sqrt{X}}{2}  S^{-2} e^+   \delta \rho  \\ 
  &\delta \omega = \frac{1}{4 \sqrt{X}} \left[ \left( 2 X^+S^- - X^-S^+  \right) S^+e^- -  \left( 2 X^-S^+ - X^+S^-
   \right) S^-e^+      
   \right] \delta \rho 
 \\ \notag \\
  &\delta X^+=  \frac{\sqrt{X}}{2} S^{+2}  X^- \delta \rho  \\
  &\delta X^-= - \frac{\sqrt{X}}{2} S^{-2} X^+ \delta \rho  \\
  \notag \\
  &\delta S=  \frac{1}{\sqrt{X}} (X^-S^+ - X^+S^- ) \delta \rho 
\end{align}
\section{Conclusions and Outlook}
\label{ConO}
The universal absolute (in space and time) conservation law for all covariant
2d theories, including interactions with matter has been shown to be related
to a Noether symmetry. Such a global symmetry had been identified before only
for the special model \cite{Kat86} in Ref. \cite{KWid94} in the absence of
matter interactions. As shown in subsection \ref{Matl} this result can be generalized
easily to all matterless models. It refers to the geometric variables (zweibeine
and spinconnection) only and may be called ``geometric'' symmetry.

When interactions with fermions and scalars are present, the ``geometric''
symmetry by itself no longer provided the complete conservation law, valid also
in that case. In addition, another (``matter'') symmetry through the conservation
of a one-form current yields a necessary integrability condition. We clarify
this unusual situation within a simplified toy-model (subsection \ref{SToy}). The symmetry
for the general 2d model, including nonminimally coupled scalars, is given in
subsection \ref{Sysf}.

Because of its importance the special case of spherically reduced gravity is
set out in somewhat more detail in section \ref{CSRG}. Here no exact solution exists.
Nevertheless we expect that the general formalism will permit appropriate gauge
choices suitable for numerical studies in which the universal conservation law
is used to fix the overall integration constant $C=C^{(g)} + C^{(m)}$
where $C^{(g)}$
 and $C^{(m)}$ are contributions from the ``geometric''
and ``matter'' parts, respectively. However, neither of these quantities may
be identified with a variable mass of the black hole in a corresponding ansatz
for the metric.

It should be stressed that in other approaches to covariant models in
two variables  the overall constant $C$ is hidden; it only appears
clearly in the first order approach advocated here, especially when
interactions with matter are taken into account.

In this connection it will be desirable also to find the ``large'' symmetry
oparations corresponding to the infinitesimal ones discussed here. They hopefully
could connect sets of numerical solutions.

All considerations of the present paper stay at the classical level. It has
become increasingly clear in the last years that even the existence of a classical
solution (with matter interactions) does not guarantee its quantum extension.
What can be done is to integrate out (exactly) the geometric variables in the
matterless case \cite{WKS92} and to treat matter as a perturbation \cite{KLV97}.
Because the universal conservation law has a quantum counterpart \cite{KLV98}
the peculiar symmetry discussed here should also be valid in the quantum case,
expressed as a Ward identity for vertex functions of 2d gravity with matter.

\section*{Acknowledgments}
We thank M. Ertl, H. Liebl and M. Nikbaht-Tehrani for helpfull discussions.
One of the authors (W.K.) has benefitted from a discussion with S. Deser. The
support of the Austrian Science Foundation (Fonds zur F\"orderung der wissenschaftlichen
Forschung), Project P 12.815-TPH is gratefully acknowledged.

\end{document}